\documentclass{article}
\usepackage[T1]{fontenc}
\usepackage{graphicx}

\usepackage{cite}
\usepackage{amsmath,amssymb,amsfonts}
\usepackage{algorithmic}
\usepackage{textcomp}
\usepackage{xcolor}
\usepackage{amssymb}
\usepackage{pifont}
\usepackage{threeparttable}
\usepackage{flushend}
\usepackage{listings} 
\usepackage{caption}
\usepackage{multirow}
\usepackage{hyperref}
\usepackage{url}
\usepackage{cleveref}
\usepackage[a4paper,
    top=60pt,
    bottom=72pt,
    left=72pt,
    right=72pt,
    headsep=24pt]{geometry}
\usepackage[misc,geometry]{ifsym}
\usepackage{authblk}       %
\usepackage{orcidlink}     %

\begin{document}
\title{Architectural Constraints Alignment in AI-assisted, Platform-Based Service Development\thanks{This work was conducted as part of a university-industry collaboration between the TU Berlin ISE department and a large German software company. The proposed framework was evaluated against deployment workflows used within the company.}}

\author{\small
    Julius Irion$^{1}$, Moritz Leugers$^{1}$, Paul Hartwig$^{1}$, Simon Kling$^{1}$,  Tachmyrat Annayev$^{1}$, Alexander Schwind$^{1}$,  Maria C. Borges$^{2}$, Sebastian Werner$^{2}$}

\affil[1]{Technische Universität Berlin, Germany \\
    \texttt{\{irion, m.leugers, paul.l.hartwig, s.kling, annayev, alexander.schwind\}@campus.tu-berlin.de}}
\affil[2]{Information Systems Engineering, Technische Universität Berlin, Germany\newline \texttt{\{mb, sw\}@ise.tu-berlin.de}}
\date{}
\maketitle              %

\begin{abstract}
AI-assisted development tools enable rapid prototyping of services but often lack awareness of architectural constraints, infrastructure dependencies, and organizational standards required in production environments. 
Consequently, generated artifacts may exhibit brittle behavior and limited deployability. 
We propose a retrieval-augmented scaffolding approach that combines platform-based code generation with agentic clarification loops to expose and resolve architectural constraint ambiguities. 
By combining template retrieval with structured interaction, the method embeds production-relevant considerations during service scaffolding.
Evaluation indicates improved architectural consistency and deployability compared to general-purpose AI code generation workflows, suggesting that constraint-aware retrieval is essential for aligning AI-assisted service development with production software engineering practices.

\end{abstract}
\section{Introduction}
The rise of Large Language Models (LLM) and AI-driven IDEs in recent years has disrupted the traditional software development process.
According to the 2025 Stack Overflow Developer Survey \cite{StackOverflowAI2025}, 84\% of respondents self-report they are using or planning on using AI tools during development.
Additionally, 51\% of professional developers report daily usage of AI tools.
This development has led to the emergence of an AI-first approach to software development often labeled as “vibe coding'' which is driven by iterative prompting.

While this approach allows for fast service prototyping, it relies heavily on a trial-and-error approach in order to generate deployable code. Consequently, this manifests itself in a cycle of prompting, the software producing errors, and re-prompting until an, oftentimes preliminary, error-free state is reached. This is not only time-consuming but also drives resource and token usage, unnecessarily increasing development costs. 

However, the limitations of vibe coding extend beyond inefficiency and also reflect a fundamental architectural mismatch. 
Software architecture demands context-specific decisions that account for organizational constraints and domain-specific knowledge \cite{ozkaya_llm4sa}.
Writing code for production environments is inherently complex, with numerous architectural considerations and subtle service dependencies that must be correctly managed: adapting to existing infrastructure services, integration with CI/CD pipelines, security policies, and alignment with established architectural patterns \cite{rag_platformknowledge}. LLMs, which rely on the ``wisdom of the masses'', lack access to this organization-specific context \cite{ozkaya_llm4sa} and consequently produce brittle deployment artifacts with inconsistent, non-deterministic behavior. Production
environments demand the opposite: \textit{consistent}, \textit{deployable} 
artifacts that satisfy organizational architectural constraints.

In this paper, we propose \textbf{a RAG-based template retrieval approach that uses agentic clarification loops} to bridge LLM-driven generation with platform engineering rigor.
Our approach leverages organizational platform knowledge (encoded in architectural templates\footnote{\url{https://backstage.io/docs/features/software-templates/writing-templates}}) as a retrieval corpus to ensure that production configurations and architectural constraints are considered when scaffolding new services. The agentic clarification loops repeatedly engage the user to best match the user's intent to the available service templates and improve the retrieval chances.
To provide initial evidence for the viability of our approach, we present a preliminary user study comparing unstructured vibe coding to RAG-enhanced platform engineering. We evaluated both approaches across three dimensions: deployment success rate, resource usage, and developer experience. Our early results suggest that the proposed approach shows promise compared to general-purpose AI-assisted service development, by ensuring that generated artifacts remain compatible with existing deployment environments, while reducing developer frustration when using AI-assistants.

The remainder of this paper is structured as follows: First, we provide background and related work in section ~\ref{sec:rw}, then introduce our RAG-based approach in section ~\ref{sec:approach}.
In section ~\ref{sec:eval}, we evaluate our RAG-based system against unstructured vibe coding.
Finally, we discuss the limitations of our study in \cref{sec:limitations} and conclude in \cref{sec:conclusion}.

\section{Background and Related Work}
In the following, we show how our tool builds upon related research and industry efforts in \textit{AI-assisted software development} and \textit{platform engineering.} 
\label{sec:rw}
\subsection{AI-assisted Software Development}
In recent years, we've seen large language models (LLMs) emerge and gain widespread adoption across the software development landscape. 
LLMs and AI-agents and are widely used as programming assistants, most notably through tools such as GitHub Copilot or Cursor. Empirical studies emphasize their usefulness in code generation \cite{llm4code_empiricalstudy_2024}, but also show that AI-generated suggestions often contain logical, syntactic, or semantic errors and frequently require manual effort to compile and behave as intended \cite{10.1145/3524842.3528470,10.1145/3715108}.

Complementing this line of work, several recent studies have explored LLM capabilities for generating architectural artifacts. Jahi\'{c} and Sami \cite{llmsoftwarearchitecturedesign} examine the state of practice by tasking LLMs with producing design and deployment diagrams, though practitioners expressed low confidence in the results. Arun et al. \cite{icsa_serverlessgeneration_2025} explore LLM-based service generation, but limit their approach to serverless deployments. Even within this constrained scope, the generated architectural artifacts were not fully functional.

Ivers and Ozkaya \cite{ozkaya_llm4sa} attribute these limitations to the inherent nature of LLMs. Architectural design requires context-specific decisions, but LLMs rely on the ``wisdom of the masses'', making them poorly suited for tasks that depend on unique organizational constraints and domain knowledge.

To address these limitations, Retrieval-Augmented Generation (RAG) approaches have recently been suggested. RAG systems enhance the generative process by grounding LLM outputs in external domain-specific knowledge sources. The researchers in \cite{rag-enhancedarchicturaldecisions_2024} apply RAG for architectural decision support, though their work focuses solely on design recommendations without extending to service generation or deployment. Pesl et al. \cite{pesl_ragservicediscovery} study RAG-based approaches for service discovery but not service generation. The idea of incorporating platform knowledge in service generation through RAG-based approaches was first proposed in \cite{rag_platformknowledge}, but this work remains purely conceptual without any implementation or validation.

\subsection{Platform Engineering}
In recent years, software architecture and engineering have been complemented by the emergence of \textit{platform engineering}. By creating toolchains and reusable services for developers, platform engineering aims to boost developer productivity, ensure operational consistency \cite{platformengineeringbook} and improve reliability \cite{srebook_platformengineering}.
\textit{Internal Developer Platforms (IDPs)} have emerged as a concrete manifestation of platform engineering principles. They encode organizational constraints and architectural decisions into reusable artifacts, providing a consistent foundation for recurrent development work. Backstage\cite{spotifybackstage,niemen2020goldenpaths} is a prominent open-source example of an IDP, originally developed at Spotify and now widely adopted across the industry.

In software engineering, we refer to \textit{scaffolding} as the automated generation of project structures and boilerplate code. When combined with IDPs and architectural templates, service scaffolding extends beyond basic code generation to include pre-configured organizational policies, security controls, and integration points for existing CI/CD pipelines and infrastructure.

\section{Approach}
\label{sec:approach}
\begin{figure}[t!]
	\centering
	\includegraphics[width=0.8\columnwidth]{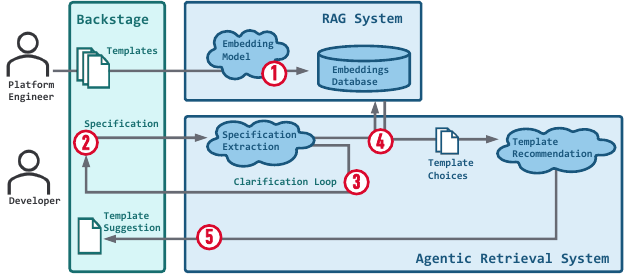}
	\caption{System Architecture and Workflow: 1. Template Ingestion 2. Conversational Specification 3. Clarification  4. Retrieval of Templates 5. Template Suggestion based on Specification}
	\label{fig:system-architecture}
\end{figure}

Our approach addresses the aforementioned shortcomings of vibe coding for service engineering by shifting from the AI-assisted generation of code to the structured RAG-based retrieval of pre-approved service templates.
We implement this as an Agentic Retrieval System integrated directly into an Internal Developer Platform (IDP), specifically utilizing Backstage\footnote{\url{https://backstage.io}}.
By grounding code generation in pre-approved architectural templates, we ensure that scaffolded services satisfy organizational constraints from the get-go.

The proposed workflow, visible in Figure \ref{fig:system-architecture}, is comprised of five distinct phases: template ingestion (only necessary once, performed by the platform engineer), requirement specification, clarification loop, retrieval, and template suggestion to the user.
The basis of our retrieval consists of a catalog of available service templates, which encompass boilerplate code, standard configurations, CI/CD pipelines, and security policies.
These artifacts are embedded using the \texttt{all-MiniLM-L6-v2} embedding model~\cite{all_minilm_l6_v2, reimers-gurevych-2019-sentence, wang2020}, which comes prepackaged with \texttt{Chroma}~\footnote{\url{https://github.com/chroma-core/chroma}}, where the templates are subsequently stored.
This results in a consolidated knowledge base for our retrieval system.

A core challenge in bridging the gap between natural language user input and structured architectural templates is the handling of ambiguity.
For this purpose, we employ an \textit{agentic clarification loop}, powered by OpenAI's \texttt{GPT4o-mini} LLM~\cite{openai2024gpt4osystemcard}. 
In order to avoid reliance on insufficient prompt input by the user, the LLM acts as a virtual software architect, repeatedly engaging the user.
This enables the effective matching of a user's intent to the available service templates and, therefore, effective retrieval, through a natural conversation with the user inside the IDP.

\begin{figure}[t]
    \centering
    \resizebox{0.95\columnwidth}{!}{
    \begin{minipage}{0.95\columnwidth}
        \hrule
        \vspace{0.5em}

        \small
        \textsf{\textbf{User:}} I need a template for a Node.js microservice
        \vspace{0.5em}

        \textsf{\textbf{LLM:}} Whats the purpose of your microservice? 
        \vspace{0.5em}

        \textsf{\textbf{User:}} It's for a product catalog connecting to our shop frontend.
        \vspace{0.5em}
        
        \textsf{\textbf{LLM:}} What type of database would you like to use? MongoDB for flexible, document-based storage, or PostgreSQL for structured, relational data?
        \vspace{0.5em}
        
        \textsf{\textbf{User:}} PostgreSQL please
        \vspace{0.5em}

        \textsf{\textbf{LLM:} }How would you like to expose your service? REST or gRPC?
        \vspace{0.5em}
        
        \textsf{\textbf{User:} }REST
        \vspace{0.5em}

        \textsf{\textbf{LLM:} }Recommended template: \texttt{node-express-postgres.yaml}

        \vspace{0.5em}
        \hrule
    \end{minipage}
    }
    \caption{Example interaction of the agentic clarification loop.}
    \label{fig:agentic_loop}
\end{figure}

Effective template retrieval requires specific information from the user.
Therefore, we define a list of requirements, including the service's purpose, the desired tech stack, and CI/CD requirements, which need to be satisfied for retrieval.
As illustrated in Figure \ref{fig:agentic_loop}, the LLM will repeatedly ask the user clarifying questions until all requirements are satisfied.

Additionally, the system is designed to handle uncertainty to accommodate non-expert users.
If the user is unable to give a specific constraint and responds with ``not sure'' or an equivalent answer, the agent disregards the strict requirement and infers the appropriate template from the remaining context.
Thereby, the retrieval still functions, even if a user is not technically proficient enough to provide all specifications.

Once the clarification loop completes and a sufficient base of information is extracted, the gathered requirements are subsequently embedded to enable querying of the template catalog.
For this purpose, we employ semantic vector search in order to retrieve the closest matching service template from the vector database.
The resulting recommendation is then suggested to the user in a final answer recommending it as the best option for the planned project.

Our approach ensures that even vague or minimal user input can be resolved to a fully functioning and deployable service scaffold.
Unlike AI-powered IDEs that merely suggest code at a granular level, our system delivers a complete architectural template that is guaranteed to work in the available environment.

\section{Evaluation}
\label{sec:eval}
To validate our claims that RAG-based service template retrieval addresses the fundamental limitations of vibe coding for scaffolding, we designed a comparative study evaluating both approaches across three critical dimensions. First, we measure \textit{deployment success rate}. This is the core indicator of whether generated artifacts actually work in production environments. Second, we track \textit{resource usage} to quantify the iterative trial-and-error cycles that characterize vibe coding and drive up development costs. Third, we assess \textit{developer experience} to understand the cognitive and practical burden each approach imposes on users. Together, these metrics capture whether our system delivers on its promise: deployable service artifacts with lower overhead and reduced developer friction.

\subsection{Methodology}

To evaluate our RAG-based system, we repeatedly tasked it to pick a correct ground-truth template among multiple close ones. Concretely, it was prompted to return a fitting template for a service with frontend and server-side rendering (SSR), that uses Postgres, and must support authentication. Correctly choosing a template with SSR was our distinguishing factor. Next to the ground truth, we used 20 other templates that deviated slightly from it, e.g., by using single-page applications instead of SSR. 
The experiment was run 10 times with randomized phrasing, but all similar in complexity. Logged metrics per run are: success (selection of the correct template), number of turns/questions asked by the agentic loop and input/output tokens. This way, we can assess how our RAG based approach performs in our quality metrics in a scenario where the correct decision is non-trivial.

To test vibe coding on the designated metrics, we devised a user study. The task we gave to testers was the following: 
\begin{quote}
\footnotesize\textsf{\textit{The software you are supposed to build is scaffolding for an Angular app with NX-Workspace, automated CI/CD pipelines and Kubernetes deployment capabilities. The CI/CD pipelines shall handle automated test, build and deployment and run on every commit.}}
\end{quote} 
A total of seven people participated in the study. All were students, doctoral candidates or post-docs recruited through the university department with which we conduct our research. The same advisor controlled the study for all executions, each of which took approx. one hour. All testers worked on the same computer, ensuring an identical setup \footnote{\url{https://github.com/paullhartwig/acp_data_and_evaluation}}. 
We supplied testers with an instance of Visual Studio Code and GitHub Copilot, powered by \texttt{GPT-5-mini}. Target for deployment was an existing Kubernetes cluster with a readily deployed GitLab runner to execute CI/CD pipelines. Each tester got their own branch in a specifically created GitLab repository. Each branch was prepared by us with deployment instructions for the AI assisted IDE that gave all necessary information about deployment, such as the cluster to target, the container registry and credentials to use.

Testers were also asked about their prior experience with CI/CD pipelines, Kubernetes, Angular, and NX Workspace, as well as how frequently they use AI-assisted IDEs. To assess whether architecting the software was successful, we had them report on the seven quality gates listed in \cref{table:vibe-results}.

The time testers took was recorded and capped at 45 minutes, after which the study was stopped. Afterwards, the prompt logs were exported from Visual Studio Code in order to calculate the token usage. We finally asked each person to rate their experience on the Likert scale and describe it in free text.

\subsection{Results for the Success Rate}

Our RAG system returned the correct template in all 10 runs with randomized phrasing. This results in a success rate of 100\%. In contrast, the results for the different quality gates in the user study on vibe coding can be seen in \Cref{table:vibe-results}.

\begin{table}
    \caption{Vibe Coding User Study Results}
    \label{table:vibe-results}
    \centering
\resizebox{\columnwidth}{!}{
\begin{tabular}{|l|ccccccc|}
\hline
\multirow{2}{*}{Quality Gate} & \multicolumn{7}{c|}{Tester} \\ \cline{2-8} 
 & \multicolumn{1}{c|}{\textbf{01}} & \multicolumn{1}{c|}{\textbf{02}} & \multicolumn{1}{c|}{\textbf{03}} & \multicolumn{1}{c|}{\textbf{04}} & \multicolumn{1}{c|}{\textbf{05}} & \multicolumn{1}{c|}{\textbf{06}} & \textbf{07} \\ \hline
 CI/CD pipelines run on every commit. & \multicolumn{1}{c|}{\ding{55}} & \multicolumn{1}{c|}{\ding{51}} & \multicolumn{1}{c|}{\ding{51}} & \multicolumn{1}{c|}{\ding{51}} & \multicolumn{1}{c|}{\ding{51}} & \multicolumn{1}{c|}{\ding{51}} & \ding{55} \\
 The CI/CD "test"-stage runs without error. & \multicolumn{1}{c|}{\ding{55}} & \multicolumn{1}{c|}{\ding{51}} & \multicolumn{1}{c|}{\ding{51}} & \multicolumn{1}{c|}{\ding{51}} & \multicolumn{1}{c|}{\ding{55}} & \multicolumn{1}{c|}{\ding{55}} & \ding{55} \\
 Building the project in the CI/CD pipeline succeeds. & \multicolumn{1}{c|}{\ding{51}} & \multicolumn{1}{c|}{\ding{51}} & \multicolumn{1}{c|}{\ding{51}} & \multicolumn{1}{c|}{\ding{51}} & \multicolumn{1}{c|}{\ding{55}} & \multicolumn{1}{c|}{\ding{55}} & \ding{55} \\
 The Docker image is successfully pushed to the registry. & \multicolumn{1}{c|}{\ding{55}} & \multicolumn{1}{c|}{\ding{51}} & \multicolumn{1}{c|}{\ding{55}} & \multicolumn{1}{c|}{\ding{51}} & \multicolumn{1}{c|}{\ding{55}} & \multicolumn{1}{c|}{\ding{55}} & \ding{55} \\
 The CI/CD "deploy"-stage runs without error. & \multicolumn{1}{c|}{\ding{55}} & \multicolumn{1}{c|}{\ding{51}} & \multicolumn{1}{c|}{\ding{51}} & \multicolumn{1}{c|}{\ding{51}} & \multicolumn{1}{c|}{\ding{55}} & \multicolumn{1}{c|}{\ding{55}} & \ding{55} \\
 The deployment itself worked, Kubernetes Pods are in state "Running" & \multicolumn{1}{c|}{\ding{55}} & \multicolumn{1}{c|}{\ding{51}} & \multicolumn{1}{c|}{\ding{55}} & \multicolumn{1}{c|}{\ding{51}} & \multicolumn{1}{c|}{\ding{55}} & \multicolumn{1}{c|}{\ding{55}} & \ding{55} \\
 The Kubernetes Pod logs do not indicate an error. & \multicolumn{1}{c|}{\ding{55}} & \multicolumn{1}{c|}{\ding{51}} & \multicolumn{1}{c|}{\ding{55}} & \multicolumn{1}{c|}{\ding{51}} & \multicolumn{1}{c|}{\ding{55}} & \multicolumn{1}{c|}{\ding{55}} & \ding{55} \\ \hline
\end{tabular}
}
\end{table}

In total, only two testers were able to pass all quality gates, while four of seven could only pass zero to one of them. One tester was able to pass four. Unsuccessful testers mostly did not make it past the early stages of the exercise, failing to execute CI/CD pipelines at all or with the desired functionality. This results in the summarized success rates of both approaches, displayed in table \ref{table:success-rates}. 

\begin{table}
 \caption{Success Rates of Both Approaches}
    \label{table:success-rates}   
    \centering
    \begin{tabular}{|l|ccccccc|l|}
\hline
\multirow{2}{*}{} & \multicolumn{7}{c|}{Tester} & \multirow{2}{*}{\textbf{RAG}} \\ \cline{2-8}
 & \multicolumn{1}{c|}{\textbf{01}} & \multicolumn{1}{c|}{\textbf{02}} & \multicolumn{1}{c|}{\textbf{03}} & \multicolumn{1}{c|}{\textbf{04}} & \multicolumn{1}{c|}{\textbf{05}} & \multicolumn{1}{c|}{\textbf{06}} & \textbf{07} &  \\ \hline
\multicolumn{1}{|c|}{Success Rate [\%]} & \multicolumn{1}{c|}{14} & \multicolumn{1}{c|}{100} & \multicolumn{1}{c|}{57} & \multicolumn{1}{c|}{100} & \multicolumn{1}{c|}{14} & \multicolumn{1}{c|}{14} & 0 & \multicolumn{1}{c|}{100} \\ \hline
\end{tabular}
\vspace{-1em}
\end{table}

Averaging the success rate of all testers, we get 43\% for vibe coding, whereas our RAG succeeds in 100\% of the cases. This suggests that our approach is reliable for selecting the correct template even in a case where multiple similar ones exist. The success of vibe coding, on the other hand, is much less probable, as not even half of the quality gates could be achieved by the testers and only two of seven were able to generate a runnable system.

Overall, we find that the vibe coding approach was largely unsuccessful and inconsistent for service scaffolding, even though the amount of external constraints to follow was comparatively small. In production environments with extensive dependencies, the architectural mismatch between LLM generation and organizational constraints becomes even more pronounced. We therefore expect deployment success rates to deteriorate further as architectural complexity increases beyond our experimental setting. This is exactly the problem we aim to solve with RAG-enhanced, template-based platform engineering, as service templates are deployable and satisfy organizational constraints by design. 

\subsection{Resource Usage}

\begin{table}[t]
    \centering
    \caption{Resource Usage of Approaches Compared}
    \label{table:resource-consumption}
    \setlength{\tabcolsep}{4pt}
    \resizebox{0.7\columnwidth}{!}{
    \begin{tabular}{|c|cccccccc|c|}
\hline
\multicolumn{1}{|l|}{\multirow{2}{*}{Costs}} & \multicolumn{8}{c|}{Tester} & \multicolumn{1}{l|}{\multirow{2}{*}{RAG}} \\ \cline{2-9}
\multicolumn{1}{|l|}{} & \multicolumn{1}{c|}{\textbf{01}} & \multicolumn{1}{c|}{\textbf{02}} & \multicolumn{1}{c|}{\textbf{03}} & \multicolumn{1}{c|}{\textbf{04}} & \multicolumn{1}{c|}{\textbf{05}} & \multicolumn{1}{c|}{\textbf{06}} & \multicolumn{1}{c|}{\textbf{07}} & \textbf{Mean} & \multicolumn{1}{l|}{} \\ \hline
Num. Prompts & \multicolumn{1}{c|}{23} & \multicolumn{1}{c|}{26} & \multicolumn{1}{c|}{24} & \multicolumn{1}{c|}{23} & \multicolumn{1}{c|}{17} & \multicolumn{1}{c|}{27} & \multicolumn{1}{c|}{14} & 22 & 3 \\
Input tok (k) & \multicolumn{1}{c|}{846} & \multicolumn{1}{c|}{1006} & \multicolumn{1}{c|}{889} & \multicolumn{1}{c|}{2043} & \multicolumn{1}{c|}{792} & \multicolumn{1}{c|}{647} & \multicolumn{1}{c|}{363} & 941 & 3.2 \\
Output tok (k) & \multicolumn{1}{c|}{10.6} & \multicolumn{1}{c|}{17.9} & \multicolumn{1}{c|}{14.5} & \multicolumn{1}{c|}{10.5} & \multicolumn{1}{c|}{11.8} & \multicolumn{1}{c|}{11.9} & \multicolumn{1}{c|}{8.0} & 12.1 & 0.26 \\
Cost \$ & \multicolumn{1}{c|}{0.23} & \multicolumn{1}{c|}{0.29} & \multicolumn{1}{c|}{0.25} & \multicolumn{1}{c|}{0.53} & \multicolumn{1}{c|}{0.22} & \multicolumn{1}{c|}{0.19} & \multicolumn{1}{c|}{0.11} & 0.26 & 0.001 \\ \hline
\end{tabular}
    }
    \begin{tablenotes}
        \centering
        \footnotesize
        \item[a] RAG values are medians over 10 runs. Cost: \$0.25/M input, \$2/M output
    \end{tablenotes}
\end{table}

The results regarding resource usage can be seen in \Cref{table:resource-consumption}. It can be seen that the RAG approach proves to be significantly more efficient than vibe coding. On the one hand, the number of prompts users exchanged with the RAG is significantly lower than that with the AI-assisted IDE. This results in less time overall spent interacting with the system. Correlating with the smaller number of prompts, the amount of tokens used by the RAG undercuts vibe coding by a factor of more than 100. In extreme cases, testers used more than two million tokens for generating a working service (see tester 04), whereas the RAG system only needed a median of about 3000 tokens. The factor of 100 for token usage in turn exceeds the factor of about ten for the number of prompts dramatically, which can be explained by the amount of context passed to the AI assistant being much larger. It needs entire files as context to work, whereas our RAG-based system merely works with small user prompts and (embeddings of) service templates. This, in turn, corresponds to a far higher cost of vibe coding as opposed to our RAG system during use.

\subsection{Developer Experience}

\Cref{table:comparison-single-col} shows the results for the developer experience testing, including the number of prompts needed, the success rate, and the rating of the overall experience, for both the RAG approach and our user study.

\begin{table}[h]
    \caption{Comparison of Vibe Coding vs. RAG Approach in Amount of Prompts needed, Success Rate and Overall Experience}
    \label{table:comparison-single-col}
    \centering
    \begin{tabular}{|c|c|c|c|c|}
    \hline
    \textbf{Tester} & \textbf{Prompts} & \textbf{Duration} & \textbf{Success Rate} & \textbf{Experience} \\
    \hline
    01 & 23 & 45 mins. & 0.14 & Bad \\
    02 & 26 & 45 mins. & 1 & Good \\
    03 & 24 & 46 mins. & 0.57 & Fair \\
    04 & 23 & 44 mins. & 1 & Good \\
    05 & 17 & 45 mins. & 0.14 & Bad \\
    06 & 27 & 46 mins. & 0.14 & Poor \\
    07 & 14 & 45 mins. & 0 & Poor \\
    \hline
    \textbf{RAG (med.)} & \textbf{3} & \textbf{\textless 5 mins.} & \textbf{1} & \textbf{n.A.} \\
    \hline
    \end{tabular}
    \vspace{-1em}
\end{table}

Our results reveal a contrast in developer experience between  RAG retrieval and the vibe coding approach. While our approach of combining RAG with predefined service templates intends to reduce cognitive load, unstructured generation through vibe coding seems to intensify it.

It can be seen that the quality of the user experience correlates with passing quality gates. Users who were unable to create deployable systems rated their participation as ``bad'' or ``poor''. In contrast, satisfying the quality gates resulted in a good experience, even when the user lacked deep technical understanding. As an example, tester 04 reported surprise at succeeding without prior knowledge.

According to the qualitative user feedback, unsuccessful users report feelings of being ``helpless'' or ``at the whims of AI'' when the assistant built a non-functional system and they were not able to fix it due to a lack of understanding the generated code. Others were satisfied for creating a basic outline of the code structure, but then got stuck when the assistant was continuously unable to resolve a specific issue. 

Furthermore,  tool expertise appeared not to improve our testers ability to create a deployable system. The expertise of testers across the different technologies used are shown in \cref{table:vibe-tester-experience}.

\begin{table}[h]
    \caption{Tool Expertise of Testers in Years}
    \label{table:vibe-tester-experience}
    \centering
    \begin{tabular}{|c|ccccccc|}
    \hline
    \multicolumn{1}{|l|}{\multirow{2}{*}{}} & \multicolumn{7}{c|}{Tester} \\ \cline{2-8} 
    \multicolumn{1}{|l|}{} & \multicolumn{1}{c|}{\textbf{01}} & \multicolumn{1}{c|}{\textbf{02}} & \multicolumn{1}{c|}{\textbf{03}} & \multicolumn{1}{c|}{\textbf{04}} & \multicolumn{1}{c|}{\textbf{05}} & \multicolumn{1}{c|}{\textbf{06}} & \textbf{07} \\ \hline
    Kubernetes & \multicolumn{1}{c|}{5+} & \multicolumn{1}{c|}{5+} & \multicolumn{1}{c|}{5+} & \multicolumn{1}{c|}{0-1} & \multicolumn{1}{c|}{0-1} & \multicolumn{1}{c|}{0-1} & 1-2 \\
    CI/CD & \multicolumn{1}{c|}{5+} & \multicolumn{1}{c|}{1-2} & \multicolumn{1}{c|}{1-2} & \multicolumn{1}{c|}{0-1} & \multicolumn{1}{c|}{0-1} & \multicolumn{1}{c|}{1-2} & 1-2 \\
    Angular & \multicolumn{1}{c|}{0-1} & \multicolumn{1}{c|}{0-1} & \multicolumn{1}{c|}{0-1} & \multicolumn{1}{c|}{0-1} & \multicolumn{1}{c|}{0-1} & \multicolumn{1}{c|}{0-1} & 1-2 \\
    NX-Workspace & \multicolumn{1}{c|}{0-1} & \multicolumn{1}{c|}{0-1} & \multicolumn{1}{c|}{0-1} & \multicolumn{1}{c|}{0-1} & \multicolumn{1}{c|}{0-1} & \multicolumn{1}{c|}{0-1} & 0-1 \\
    AI-Development & \multicolumn{1}{c|}{F} & \multicolumn{1}{c|}{A} & \multicolumn{1}{c|}{A} & \multicolumn{1}{c|}{F} & \multicolumn{1}{c|}{F} & \multicolumn{1}{c|}{F} & N \\ \hline
    \end{tabular}

    \begin{tablenotes}
        \centering
        \footnotesize
        \item[a] A = all the time, F = frequently,  N = never
    \end{tablenotes}
    \vspace{-1em}
\end{table}

Tester 01, despite 5+ years of Kubernetes experience, failed all quality gates and rated the experience as `bad'. This suggests that for experts, the negative experience stems from the friction between their correct mental model and the AI's tendency to architect with wrong assumptions or false versions, which was also described as irritating by our testers. However, tester 04 was successful despite not reporting experience with any of the technologies, indicating, if at all, a negative correlation between technological expertise and success with vibe coding. 

Generally, it can also be seen that all testers exhausted the 45-minute time-cap. Furthermore, the number of prompts with which they attempted to architect a working system greatly exceeds what was necessary with the RAG approach. 

In contrast, our RAG-based approach effectively mitigates the psychological stressors that building services with the AI assistant has shown to cause by enforcing a `correct-by-design' paradigm. Retrieving pre-validated templates rather than generating code from scratch eliminates the risk of subtle architectural hallucinations. Users are not required to debug complex infrastructure configurations generated by an AI. Instead they are provided instantly with service scaffolds that inherently satisfy organizational constraints and can be deployed with the click of a button.

\section{Limitations}
\label{sec:limitations}
Although our results demonstrate a clear advantage for the RAG-based template approach, limitations regarding the experimental design must be considered when interpreting our findings.
A primary limitation of our user study is the reliance on a small sample of seven testers. Furthermore, since our participants were recruited from an academic environment, they may lack the specific knowledge possessed by full-time engineers. Therefore, the high failure rates observed in the vibe coding group might differ in an enterprise setting where domain expertise is higher and people are more well-versed with the tools in use. Additionally, we only tested with a single task, which could be expanded to more cases in a more thorough analysis. Also, the quality gates could be expanded; we only verified the Kubernetes pod logs to be error-free but did not actually use the AI generated service or verify correct functionality. Lastly, we used the same GitHub login for each test iteration, so it cannot be ruled-out that the system learned from each iteration. Our tests results, however, did not suggest this.

Our RAG approach came out favorably in the comparison, however the evaluation focuses on the transactional cost of deployment by measuring the tokens and time required for a developer to build a service. This metric does not show the operational cost required to engineer, validate and maintain the underlying templates. Unlike vibe coding, which offers an environment capable of attempting any task immediately, the utility of the RAG system is strictly conditional on the existence of a relevant, pre-approved template. The efficiency gains reported in our study reflect a shift in responsibility: from the application developer to the platform engineer.

Lastly, it should be noted that the ecosystem around developer platforms is evolving rapidly. During the development of this work, Backstage introduced support for the Model Context Protocol. MCP enables AI systems to interact with platform tools and services through standardized interfaces, potentially allowing AI assistants to access platform capabilities, templates, and documentation directly. As this feature became available only late in our development process, we were not able to evaluate it within the scope of this study, but it represents a promising avenue for future work.

\section{Conclusion}
\label{sec:conclusion}
In this work, we presented a comparative evaluation of two approaches to service scaffolding: unstructured, prompt-driven vibe coding and a structured, RAG-enhanced template retrieval system. Our RAG-based system has been shown to beat vibe coding in all recorded metrics of deployment success rate, resource consumption, and overall developer experience for scaffolding new services.

While testers were able to create a runnable system through vibe coding only in two of seven cases, our RAG-based approach recommended the correct, inherently deployable template 100\% of the time while also requiring far fewer interactions with the user and time to be spent with it. Generally, testers reported a great deal of psychological stress and a rather bad experience while interacting with the AI-assisted IDE, especially in cases where it did not work as intended. This problem is alleviated by our system through a superior success rate and deployability-by-design of the result. At the same time, it is more resource-efficient by a factor exceeding 100x.

While LLMs and agentic code generation have lowered the barrier to service development, it can be seen that these tools are currently unable to handle intricacies, organizational constraints and non-trivial requirements of production-grade service architectures. They have shown here a tendency to trap developers in frustrating ``trial end error" phases that lead to high token usage and feelings of helplessness when AI generation fails. 

In contrast, our RAG-based approach of recommending fitting, pre-defined templates has shown to act as a guardrail, ensuring deployability of services and reducing resource usage. This, however, comes with an organizational necessity: The benefits of such a system can only be had if the organization using it possesses such pre-defined templates.
This requires an up-front investment in building a catalog of templates for service scaffolding that, as organizational constraints change and develop, also has to be maintained to avoid architectural drift. This means adopting a platform engineering approach to software architecture. However, the results of this work indicate that this is a worthwhile investment.

\bibliographystyle{splncs04}
\bibliography{refs}

\end{document}